# Controlling microgel morphology and swelling behavior by copolymerization

Domenico Truzzolillo[1*], Thomas Hellweg[2], Julian Oberdisse[1*]

[1] *Laboratoire Charles Coulomb (L2C), University of Montpellier, CNRS, 34095 Montpellier, France.*
[2] *Department of Physical and Biophysical Chemistry, Bielefeld University, Universitätsstr. 25, 33615 Bielefeld, Germany*

*Authors for correspondence : domenico.truzzolillo@umontpellier.fr, julian.oberdisse@umontpellier.fr*

**Abstract**

The thermosensitive behavior of microgel particles suspended in solvents, i.e. their temperature-dependent swelling properties, has triggered ongoing interest in industry and academia over the past forty years. The most-studied polymer is poly(N-isopropylacrylamide) - PNIPAM -, where the volume phase transition temperature is well known to depend on the detailed molecular architecture of the monomers. In this article, we focus on publications mostly of the past five years in chemical synthesis, aiming at shifting or controlling the volume phase transition temperature (VPTT) of such polymers by copolymerization of a main monomer – often from the PNIPAM family – with either monomers of different hydrophobicity, or with ones bearing ionizable groups. In some cases, hydrophobicity may be modulated by light as external switching parameter, whereas ionic strength or pH may act on the thermosensitivity of the microgels containing charged groups. Due to either differences in reactivity, or specific synthesis routes, particular microgel morphologies, such as molecular gradient, core-shell, interpenetrated, or patchy (multi-lobular) structures may be generated. They may give rise to spatial modulations of thermosensitivity within particles and are highlighted in this review. Our short overview shows that multiple external control of VPTT and morphology is commonly achieved nowadays.

# 1 Introduction

Microgel particles are finite-sized, typically sub-micron crosslinked polymer networks suspended and solvated in a solvent, usually water [1] **. The most-studied chemistry is based on acrylamides, such as poly(*N*-isopropylacrylamide), PNIPAM [2]. Such particles are internally crosslinked, often with N,N-methylenbisacrylamide (BIS) [3–5]. A range of different chemistries, some of which will be discussed in this review, is available. The most striking property of microgels is their ability to change their solvation state—and consequently their size—in response to temperature, making them key players in the vibrant field of soft nanotechnology [6] and, more generally, of soft condensed matter physics and chemistry [7–10]. PNIPAM, for instance, undergoes a sudden deswelling, called the volume phase transition (VPT), upon heating above approximately 32°C. This VPT-temperature (VPTT) is close to the physiological temperature, and this lucky coincidence has triggered many studies in the field of bio-medical applications. As we will not specifically highlight such applications here, the reader is referred to the reviews by Das et al. on (bio)sensing applications of PNIPAM [11], and to the review by Zhang and Gao for medical applications of microgels in general, with focus on synthesis techniques [12]. Many groups have turned towards the study of more and more complex syntheses, aiming at new functionalities [2,13–19]. They are all based on the fundamental thermoresponsiveness of the microgel particles, and adjusting this property is essential for adapting to specific conditions. Therefore, this review is deliberately restricted to the modification of the swelling behavior by copolymerization.

The VPTT can be measured by techniques such as turbidimetry or dynamic light scattering (DLS, also called photon correlation spectroscopy, PCS), the latter providing the hydrodynamic radius $R_H$ as a function of temperature in case of dilute samples. The resulting $R_H(T)$ curves are called swelling curves, and they can be modeled by relatively simple empiric functions [20], while the general, though debated, thermodynamic framework is provided by the Flory-Rehner theory [21–23]. Other techniques sensitive to particle size and collective structure or dynamics include rheology [24–27] and scattering techniques (SLS, PCS, SAXS, SANS) [5,9,28], while changes in solvation, segmental dynamics or microgel interactions and adsorption can also be followed by NMR [29,30], tensiometry [31,32] and surface force measurements [33,34]. It is noteworthy that the combination of static and dynamic scattering provides a comparison of $R_H$ to the radius of gyration $R_g$, and thereby an indication of internal mass distribution within microgels [20,28]. More evolved analyses of microgel morphology can be achieved by AFM [35], or specific modelling of SAXS or SANS data [5,36–39], possibly with deuteration for highlighting substructures or specific monomers.

It is worth recalling that for more dense samples standard PCS measurements should not be used since multiple scattering can occur and apparently enhance the dynamics. In this case the use of more advanced 3D cross correlation light scattering techniques is preferable [40,41].

The underlying cause of thermosensitivity of microgels is the temperature-dependent hydrophobic effect. The phase diagram of certain monomers in water displays a decrease in solubility at higher temperatures, above the lower critical solution temperature (LCST). For polymer molecules made of such a monomer, the solvent quality decreases with temperature, and the polymer conformation changes, from highly solvated coil structures in good solvents to a collapsed state in bad solvent. If these polymer chains are the subchains of a crosslinked microgel, then this mechanism induces the temperature-dependent deswelling of the particle. Taking a more hydrophilic monomer – which may be non-ionic or bear electrostatic charges, opening the possibility of control by pH or salt – usually leads to a higher VPTT. In contrast, a more hydrophobic monomer has the opposite effect. Synthetic strategies therefore explore either changing the average hydrophobicity by random copolymerization, or introducing parts or blocks of different solubility, thereby forming more complex, hierarchical structures. Such strategies of controlling the swelling properties of the microgel by copolymerization are the topic of the present review.

In this article, contributions dating mostly from the past 5 years are reviewed, regrouped in five sections. The focus of the first two parts lies in the modulation of hydrophobicity by copolymerization with more or less hydrophobic molecules, inducing shifts in the VPTT, without (section 2) or with (section 3) the formation of non-trivial morphologies, such as different spatial compositional profiles or core-shell structures, interpenetrating networks, or non-spherical, patchy shapes. In section 4, the effect of copolymerization with photosensitive monomers is presented. In the last parts, finally, copolymerization with ionizable monomers is discussed (section 5), giving rise to external control of thermosensitivity by pH or ionic strength. Again, a separate section (6) highlights studies focusing on non-trivial morphologies.

## 2 Controlling transition temperature of spherical thermosensitive microgels by copolymerization with hydrophobic or hydrophilic comonomers

The general strategy of the approaches highlighted in this first section is to control the thermal response of microgel particles by copolymerizing randomly a main monomer with a comonomer of different hydrophobicity, see for instance [42], possibly bearing a second functionality besides T-dependence. We start by reviewing several articles which exploit a non-NIPAM main monomer. Vinylcaprolactam (VCL) has been the molecule of choice for several authors, due to its low toxicity and high biocompatibility. Nizardo et al. present such a VCL-based microgel with dual responsiveness due to the introduction of a more hydrophilic, non-thermoresponsive comonomer, N-methyloacrylamide (NMA) [43]. As expected, the higher hydrophilicity leads to a higher VPTT, increasing from some 25 to 35°C. Moreover, the transition is pH-sensitive in the present case. Indeed, at low pH, more H-bonding further increases the hydrophilicity, and thus the VPTT. Crosslinking by BIS (here called MBA) is also studied, showing that crosslinking brings in the necessary hydrophobicity to move the thermal response in the region of physiological temperatures.

Dieuzy et al. have published an explicit study of the effect of different crosslinking of another non-NIPAM-based microgel, a copolymer of di(ethylene glycol)methylethermethacrylate and oligo(ethylene glycol) methylethermethacrylate (OEGMA), where "oligo" refers to typically eight ethylene glycole (EG) groups [44] . The swollen state of the particles is studied by rheology, and these authors have shown that the looser, more hydrophilic OEGDA-crosslinking – a diacrylate (DA) of the same family as the main monomer – leads to more swollen particles as compared to BIS-crosslinking. One would expect that OEGDA-crosslinked particles reach interaction and thus increase in viscosity at lower nominal volume fractions, but the opposite is observed. Exploiting NMR, the authors show that the OEGDA-crosslinking induces the formation of more homogeneous particles, whereas BIS leads to denser cores and extended coronas.

A third article which we would like to present here takes the stand of proposing new thermosensitive microgel particles based on cheaper and industrially available monomers, namely acrylamide and the more hydrophobic diacetone acrylamide (DDAM) [45]. Although both monomers are not thermosensitive in the accessible temperature range, they show that the thermo-responsiveness of the resulting copolymer particles exists. Moreover, it can be adjusted between a weak swelling response at high temperature (70°C) for small amounts of DDAM, down to room temperature for higher DDAM feeds. This interesting result suggests that many more surprises in copolymerization of these compounds are yet to come.

The VPT of NIPAM-based copolymer microgels with the non-thermoresponsive comonomer N-tert-butylacrylamide (NtBAM) has been studied by Krüger et al. [46]. Interestingly, the two monomers differ only in the presence of the iso-propyl in contrast to the tert-butyl group, motivating an interpretation in terms of steric demand. As we will see in other contributions, the more hydrophobic comonomer tends to decrease the VPTT continuously and leads to a higher number of nucleation sites at high temperature in water during synthesis. The resulting particles are smaller, and both VPTT and hydrodynamic size evolutions are fitted with a model based on Flory-Rehner theory, as shown in Figure 1. Note that at high NtBAM content, the deswollen state saturates, indicating that then both monomers possess similar hydrophobicity. In another work by the Bielefeld group, with focus on secondary crosslinking for membrane formation which is not the scope of the present review, the authors show that an additional, strongly hydrophobic crosslinker, also leads to a corresponding decrease of the VPTT [47].

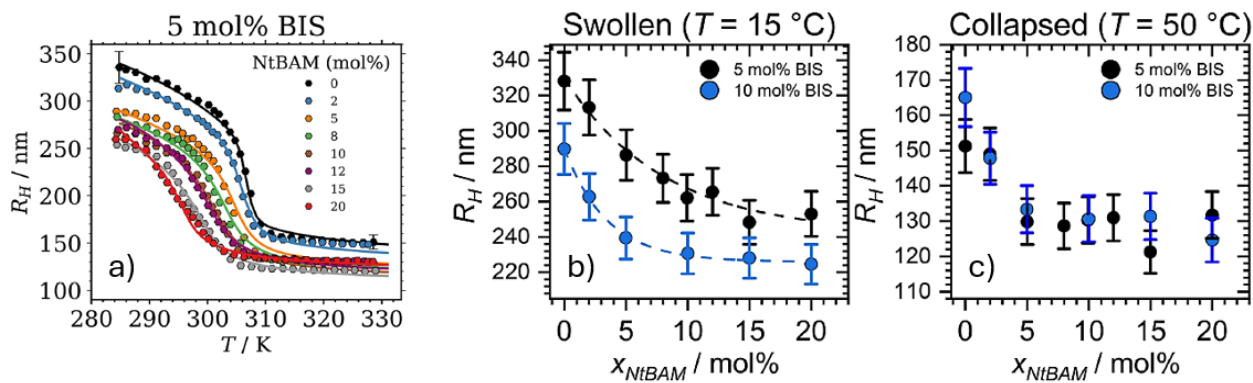

**Figure 1**: a) Swelling curves for PNIPAM-*co*-NtBAM at 5 mol% BIS crosslinker and varying NtBAM copolymer content. Solid lines are the best fit to the data of the Flory-Rehner model developed in [46] b-c) Hydrodynamic radius $R_H$ at fixed temperatures with varying NtBAM contents at 15 °C in swollen state and 50 °C in collapsed state. Adapted from [46], licensed under CC BY 4.0.

Often copolymerization is motivated by a specific function or application. Wang et al. prepared NIPMAM-based microgels copolymerized with a hydrophobic comonomer, leading to poly(N-isopropylmethacrylamideco-4-acrylamidobenzo-18-crown-6) microgels capable of selectively capturing lead ions in aqueous solution [48]. In line with our previous reasoning, the incorporation of hydrophobic compounds tends to decrease the VPTT continuously. Moreover, the VPTT is modified in presence of $Pb^{2+}$, indicating binding of the latter to the microgel. This is not the case for other ions ($K^+$, $Na^+$, $Ca^{2+}$, $Cd^{2+}$, $Ba^{2+}$, $Sr^{2+}$), showing the selectivity of the association between modified microgels and lead ions. This study is conceptually related to articles by Hohenschutz et al. [49]*, who study spontaneous binding to microgels of superchaotropic polyoxometalate ions, i.e. rather large ions with rather low charge density. These ions seem to have no effect on the VPTT but can be used to tune the total microgel swelling. Electrostatic repulsion induces larger swelling at low ion concentration, whereas the same ions at higher concentration add additional crosslinking, thereby reducing swelling.

NIPAM-based microgels can also be copolymerized with compounds which are more hydrophilic than NIPAM itself at low temperatures. Of note, carboxymethylcellulose should increase the VPTT, and convey degradability and biocompatibility to the material. With a drug delivery application in mind, Amoli et al. have copolymerized NIPAM with different contents of methylcellulose, showing that the presence of the latter increases the VPTT from the one of pure PNIPAM to roughly 40°C [50]*. They have then investigated the influence of BIS-crosslinking in phosphate buffers to imitate physiological conditions on the release of Genipin, a biomolecule extracted from fruit. Genipin possesses regulating properties in drug delivery and cell differentiation for future tissue engineering. A higher crosslinking, corresponding to 10 mol% of BIS as opposed to a lower value of 2.5 mol% was found to allow higher drug loading, but a slower Genipin release, due to the tighter network.

As a final note on modifications of the VPTT by copolymerization with neutral molecules, the importance of deuteration should not be underestimated. Indeed, many structural studies by small-angle neutron scattering (SANS) rely on the creation of artificial scattering contrast by introducing deuterated compounds. For NIPAM-based copolymers, the effect of deuteration on the swelling properties has been studied by some of us [51]. The key results to be kept in mind when designing a scattering experiment with molecules such as NIPMAM are that the VPTT of deuterated polymers is higher, and that its shift is not related in a simple manner to the number of deuterium atoms in a monomer.

## 3 Controlling thermosensitivity by copolymerization with hydrophobic or hydrophilic comonomers, with focus on microgel morphology

In the previous section we have discussed recent contributions where we believe the authors were aiming at the synthesis of spatially homogeneous microgels with shifted VPTTs obtained by random copolymerization with comonomers of different hydrophilicity. The latter property drives the different propensity of the comonomers to precipitate during synthesis in water, and, together with different chemical reactivities, it may lead to the formation of spatially inhomogeneous microgels. Such microgels may thus present new structures, with original swelling properties. Going one step further, one can even introduce different timing for the monomer feeds, thereby hoping to create new structures. The group of W. Richtering, in particular, has a strong record in creating original structures, including hollow microgels obtained through the use of sacrificial inorganic cores, with innovative loading and delivery applications in the line of view [52] – one of their articles will be reviewed here. Other groups make use of the microgel particles to grow inorganic cores. The latter may generate new properties such as plasmonic ones, thereby introducing new functions [53]. In what follows, however, we restrict ourselves to the discussion of purely organic microgels.

We start again with VPTT modifications of non-NIPAM systems, this time with focus on non-trivial morphologies. Walkowiak et al. have published their strong efforts in synthesis of patchy microgels [54]. The latter are obtained by swelling a copolymer microgel made of vinylcaprolactam and OEGMA by styrene. Depending on the amount of BIS-crosslinker in the microgel, the relaxation of the styrene-swollen structure leads to either big patches on the surface of dense microgels, or to an almost smooth surface in the opposite case.

In their contribution, Sasaoka et al. focus on an inverse miniemulsion RAFT synthesis, including hydrophilic and biological components, which is difficult in conventional high-T precipitation polymerization [55]. They designed a zwitterionic core made of the phospholipid 1-palmitoyl-2-myristoyl-sn-glycero-3-phosphocholine PMPC with a thermoresponsive POEGMA-co-MEO$_2$MA shell, the latter conveying a VPTT of 38 °C – slightly higher than the PMPC-free copolymer – to the entire particle as observed by turbidity measurements. The combination of a non-thermosensitive core with the swelling properties of a shell thus provides the opportunity of loading additional biological compounds while keeping control of, e.g., colloidal stability by temperature.

In another publication on non-NIPAM-based microgels, Sonzogni et al. explore the formation of more complex structures based on vinylcaprolactam, which is again chosen for its better biocompatibility [56]*, despite the fact that polymerised NIPAM is also found to be biocompatible in cell culture applications. On the other hand, it is not film-forming, i.e. drying suspensions of such particles does not lead to the formation of a mechanically stable, solvent-free polymer film. They therefore synthesized cores of crosslinked p(VCL) and added a shell containing the low-$T_g$ polymer, butyl acrylate, which is known to be film forming, but not thermo-responsive. The resulting microgels are patchy, the p(VCL-co-BA) lobes being capable of interconnecting into a 3-D film upon drying. These authors then checked release from films of the protein ovalbumine. Interestingly, the synthesis temperature can be used to tune the resulting film properties. For reaction temperatures above the $T_g$, the patchy shell is liquid and seems to evolve into a complete layer. It allows film formation, but then its hydrophobic barrier properties impede release of hydrophilic compounds. On the other hand, mass transfer is observed in presence of a low and controlled number of patches.

The hydrophilic surface modification of preformed microgel cores has been proposed by Gruber et al. with a view to protecting hydrophobic zones for molecular transport and delivery in biological environments [57]**. The obtained synthetic platform and its versatility is sketched in Figure 2. PEGylation, i.e. the protection by an outer PEG shell, is a technique designed to increase colloidal stability and biological half-life. Gruber et al. choose a different pathway, where the surface modification is performed with a macromolecular surfactant containing a hydrophobic moiety capable of grafting onto the growing core, and a hydrophilic one made of poly(ethylene glycol) methyl ether methacrylate. The latter carries a reactive alkyne group with one triple carbon-carbon bond for possible post-functionalization, creating a library of particles with different properties. The pentafluorophenyl methacrylate core itself is dual responsive, as it contains both pH-responsive polymer, and a cleavable disulfide crosslink, conferring redox sensitivity to the core. The authors showed by DLS that pH-dependent swelling is indeed observed, as well as reduction-induced degradation via the reduction-cleavable crosslinker in the core.

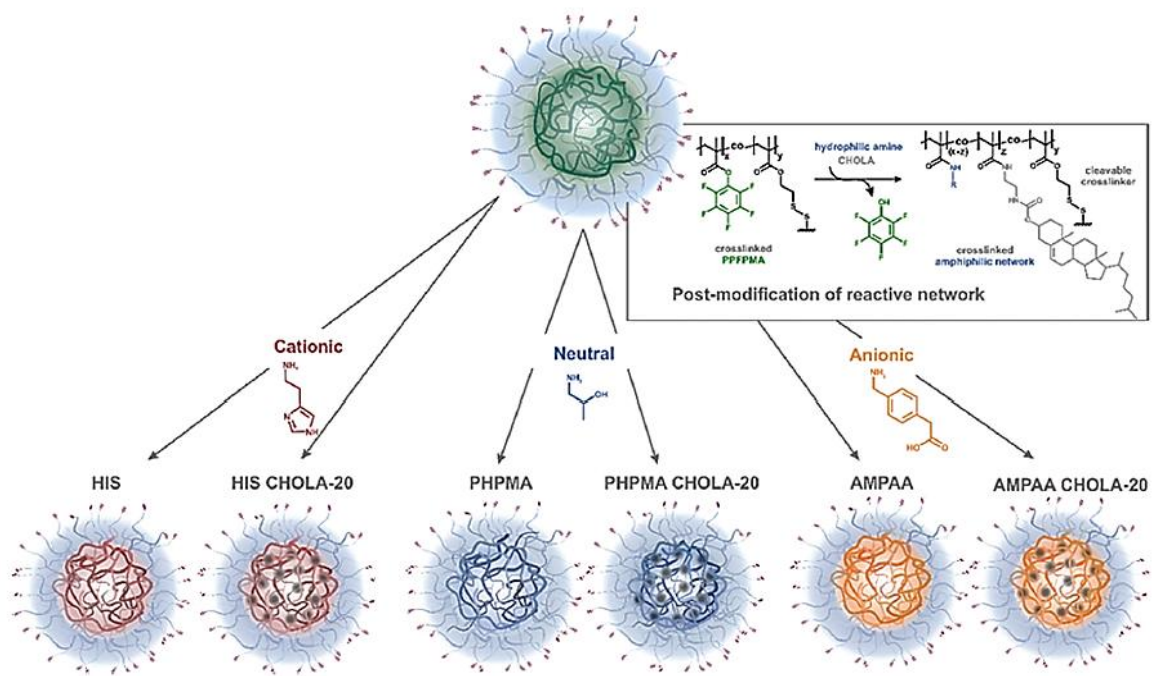

**Figure 2**: Schematic representation of the synthetic platform for preparation of functional amphiphilic microgels (crosslinker fraction 2 mol%) with a hydrophilic, reactive shell detailed in [57]**. Functionalizing the interior of reactive precursor particles with different hydrophilic moieties gives access to a library of microgels with opposing pH response (cationic, neutral and anionic microgels). Incorporation of additional 20 mol% of hydrophobic amine-functionalized cholesterol results in amphiphilic microgels with a reactive hydrophilic shell that can be used for further functionalization. Adapted from [57]**, licensed under CC BY-NC 3.0.

The rheological response of dense NIPAM-based microgels has been related to their internal structure by Bassu et al. [58]*, in a follow-up paper of a first experimental study [59] which was then further analyzed with help of numerical simulations [60]. Bassu et al. copolymerized NIPAM (ca. 66 wt%) with non-thermoresponsive poly(ethylene glycol) methyl ether methacrylate (PEGMA, ca 34 wt%), where the PEGMA is relatively short, and introduced only low quantities (1 mol%) of ethylene glycol dimethacrylate (EGDMA)-crosslinker. EGDMA is expected to react faster than BIS with PNIPAM [61]. This resulted in very soft microgel particles with small cores containing most of the crosslinker, and extended soft coronas, similar to the picture proposed by Dieuzy et al. [44]. Such particles were found to have a VPTT of ca. 36°C, i.e. the transition is shifted to slightly higher temperatures under the influence of the more hydrophilic comonomer. The monomeric density profile of the microgel particles has been measured by SANS. The authors found a smooth, star-like density profile, as in ultra-lowly crosslinked microgels, and although no contrast variation is used, they conclude from the shift of the VPTT that the PEGMA is well incorporated in the microgel, as opposed to being grafted on the microgel surface. Upon increasing concentration, the soft microgels were found to undergo deswelling, and then a sudden increase in rigidity, until finally losing the individual microgel nature, the dense suspension turning into a macroscopic hydrogel.

In a series of articles, Cors et al. have studied core-shell microgels made of different monomers of the NIPAM family, namely NNPAM, NIPMAM, and NIPAM, with VPTTs ranging between 20 and 45°C [62]. The original idea was to study the effect of a linear thermoresponse of such microgels, i.e. a continuous decrease of the hydrodynamic radius with temperature, as observed by Zeiser et al. [63]. By selectively deuterating one of the two monomers and using contrast matching in different mixtures of $H_2O$ and $D_2O$, the density profiles of the different monomers could be determined from the scattered intensities using a reverse Monte Carlo approach. This allowed showing that the observed linearity in deswelling was due to an effective interpenetration of the second, "shell"-monomer into the core. As a result, a shell at a given distance from the center of the microgel could be attributed to a VPTT corresponding to the local monomer concentrations, and the progressive collapse of these shells was thus proposed to explain the apparent continuous deswelling.

In more recent contributions, the copolymerization of NIPAM- and NIPMAM-based microgels with either hydrophilic comonomers, namely purpose-synthesized N-(2-hydroxyisopropyl)acrylamide (HIPAM) or N-(hydroxymethyl)acrylamide (HMAM), respectively, [64], or N-(1-hydroxy-2-propyl)methacrylamide (NIPMAMol) [65] was studied with the aim of increasing the VPTT. None of these molecules shows thermosensitivity in the accessible temperature range in water at ambient pressure, but incorporating such hydrophilic compounds at different contents leads to a steady increase of the VPTT. This evolution can in all cases be extrapolated to the hypothetical VPTT of the pure hydrophilic compound, and values ranging from ca. 70°C (HIPAM), to 85°C (NIPMAMol), and then almost 100°C (HMAM) have been evidenced. Unfortunately, it was not possible to synthesize such pure - but nonetheless BIS-crosslinked - microgels. These high values illustrate the strong impact of such comonomers on the VPT: the transition becomes smoother, the microgel particles grow, and the transition temperature goes up. From a fundamental point of view, this behavior was linked to the different, and possibly late or incomplete precipitation of the more hydrophilic compounds during synthesis. And, finally, concerning possible energy applications, such as security add-ons for fuel cell membranes, the VPTTs reached are a clear step in the right direction, towards some 80° or 90°C, albeit not far enough yet [14].

## 4 Controlling thermosensitivity by copolymerization with light-sensitive comonomers

Up to here, we have reviewed articles where the temperature-dependent swelling and deswelling of the microgel particles can be controlled by copolymerization aiming at modifying the hydrophobic-hydrophilic balance. In the later subsections, we will discuss articles where copolymerization introduces ions or ionizable groups. The swelling can then be modified by changing parameters such as ionic strength or pH, necessitating physico-chemical manipulations. In contrast, shining light as an external trigger is non-chemical, considerably easier to apply, to reverse, and to control both in space and time. The existence of photo-sensitive molecules, which adopt different conformations in water, of different hydrophobicity as a function of the energy of the incoming photons, is an elegant way of modulating their VPTT by light. By copolymerizing them with microgel-forming molecules, an efficient way to control the thermosensitivity of the resulting microgels has been proposed. In particular azobenzenes and spiropyrans have been studied in this context because of their reversible transition between cis- and trans-states [66].

Phua et al. have published the synthesis and swelling properties of biocompatible vinyl-caprolactam copolymerized with small amounts of 4-[(4-methacryloyloxy) phenylazo] benzenesulfonic acid [67]. This copolymerization leads to smaller particle sizes and to a lower softness of apparently more hydrophobic microgels with a VPTT lowered by several K. PCS experiments have shown that illumination with UV-light during approximately 1h further lowers the VPTT, and induces deswelling at temperatures between 15 and 30 °C, while the VPTT of the pure VCL crosslinked by BIS is ca. 33°C. This is surprising, as UV illumination triggers the transition to the more polar cis-state. A possible explanation put forward by the authors is that the internal reorganization of the charges of the sulfonic acid is modified, possibly disrupting additional H-bonds. The observed deswelling can be reversed by illumination with visible light: a remarkable sequence is shown in Figure 3. Besides providing an efficient way to modify externally microgel hydration states, this article illustrates the complexity of intramolecular interactions at play in such systems, particularly the steric hindrance due to bulky side groups.

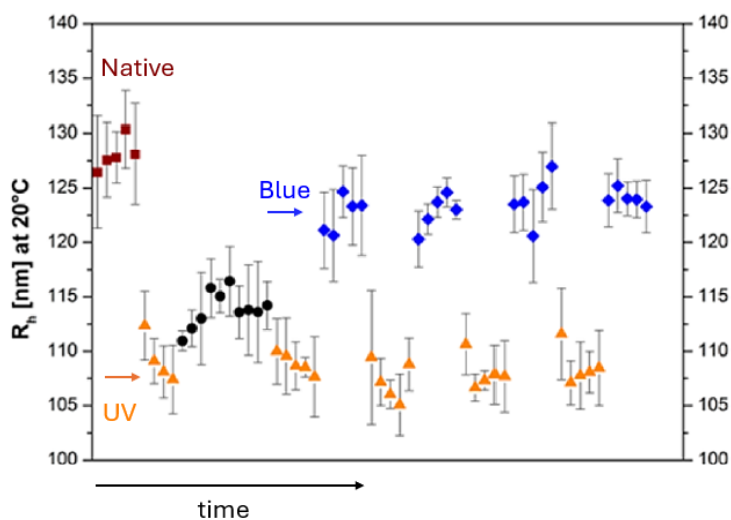

**Figure 3:** Investigation of the photoswitchability in size of P(VCL-BIS$_{1.5}$-ABSA$_{4.5}$) microgels at 20 °C, over several cycles, via PCS measurements. ■ represent the $R_H$ of the microgels in their native states, ▲ represent the $R_H$ reached with UV irradiation (λ =365 nm, 100 mW cm$^{−2}$), ● represent the $R_H$ of the particles upon removal of UV and left to relax in the dark,

and ◆ represent the $R_H$ attained with blue light irradiation (λ = 450 nm, 1.0 mW cm$^{-2}$). Each measurement point represents the mean of 6 × 10 s acquisitions in the PCS instrument. Adapted with permission from [67], Copyright 2016 American Chemical Society.

In a second contribution by Andrij Pich's group on a system also based on VCL, Hu et al. have developed a new method to synthesize light- and temperature dual-responsive microgels modified with high loading of spiropyran [68]. To overcome difficulties in direct microgel-forming copolymerization, they have opted for a copolymerization with vinylformamide, which could then be hydrolyzed and coupled to carboxyl-modified spiropyran. The resulting linear molecules show the expected shift in LCST from some 67 °C down to about 50 °C under irradiation. These linear chains have then been crosslinked into microgels containing more than 10 mol% of spiropyran in an additional step, as compared to quantities below 5% in their previous work discussed above [67]. The presence of VCL and spiropyran makes the microgels both thermo- and light-sensitive. In the dark, they achieve a VPTT of 19.5 °C, which is 5 °C lower than the spiropyran-free analogue. An impressive reversible switching in microgel size within 10 min of irradiation is evidenced, between more hydrophobic and smaller particles under light, and bigger particles in the dark. With increasing temperature, the particles deswell progressively, and the resulting size-response with light becomes less pronounced. In a follow-up paper, Hu and Pich designed and studied multiresponsive microgels via post-modification of PVCL-co-AAC microgels, where the acid groups provided additional pH-response [69]*. The authors showed that around the VPTT, the microgels could be swollen (resp. deswollen) depending on the wavelength of the incoming light. Moreover, the charge of the particles could be reversed when going from acid to basic solutions.

In addition to that, in 2016 Zhang et al. have investigated the properties of NIPAM microgels also incorporating spiropyran by copolymerization with acrylic acid which was subsequently functionalized by spiropyran [70]. The group has a long-standing expertise with the construction of etalons, which are gold layers separated by a single microgel layer, the nanometric distance between the gold layers thereby depending on microgels swelling. Consequently, the optical properties and in particular colors reflect the hydration state of the microgels. Using such etalons, the authors showed that the resulting microgels were smaller due to the replacement of the acid group, and that they exhibited response to temperature, UV-vis, pH, and to the presence of copper ions ($Cu^{2+}$).

More recently, a completely different approach to incorporation of spiropyran has been proposed by Sharma et al. [71]. They complex a photosensitive, charged amphiphile with anionic microgels in the swollen state, reaching a distribution of the surfactant inside the particle due to charge compensation. This leads to a reduction of the microgel size. Upon additional UV-light irradiation, the azobenzene part becomes more hydrophobic, and further reversible shrinking is observed.

**5 Controlling thermosensitivity of spherical microgels by copolymerization with ionizable comonomers**

Many groups have published work on copolymerization of NIPAM-based systems with charge-carrying monomers focusing on their thermosensitive swelling properties. They usually do not pay particular attention to the overall morphology, which is described as approximately spherical. The typical ionizable comonomer is a weak acid, the most common ones being acrylic or methacrylic acid. We start with an early article on this topic, which is now over twenty years old and still highly cited. Hoare and Pelton performed a comparative study of the incorporation of various acidic comonomers, in particular vinylacetic acid, acrylic or methacrylic acid [72]. The swelling of NIPAM-vinylacetic microgels is found to be more pronounced than the one of the other two systems. While the VPTT at low pH remains comparable to the one of pure NIPAM microgels, upon ionization at high pH the VPT is shifted to more than 50°C for moderate vinylacetic acid contents, and to at least 70°C for the highest contents. Obviously, the pH provides an external trigger capable of tuning the VPTT, while the feeds allow adjusting the system to a VPTT range of interest.

The same group addressed the differences in distribution of carboxylic acid functional groups in PNIPAM-based microgels, when they were directly copolymerized or functionalized post synthesis with methacrylic acid and acrylamide [73]. The way the charged comonomers are introduced, has a major impact on the comonomer distribution.

In a contribution by one of us, Elancheliyan et al. have studied NIPAM-based microgels copolymerized with acrylic acid in order to highlight the influence of the electrostatic charges on the swelling behavior [28]. As with other hydrophilic comonomers, the particle size is found to increase, the VPTT shifted to higher temperatures (by some 15 K), while the transition itself is broadened, until its complete disappearance for 5% of acrylic acid. An important aspect of this study in the framework of the present review is that the authors compare the total particle size expressed by the hydrodynamic radius to the radius of gyration which is indicative of the mass distribution inside the microgels. These two characteristics of the particles are found to decouple upon increasing charge ratio, with the internal structure of the microgels ($R_g$) collapsing due to hydrophobic interactions, while the charges still favor an extended conformation, and thus a higher $R_H$. This article thus provides evidence for the onset of the formation of a highly heterogenous morphology induced by electrostatic charges during the (two-step) deswelling, a finding which is furthermore backed up by computer simulations.

The influence of the molecular structure of microgels has been investigated by Buratti et al. [74]*. This group succeeds in comparing the swelling properties of random networks with interpenetrating ones, where pure NIPAM-cores are used as template for the polymerization of a second acrylic acid network. In good agreement with the findings of the previously discussed article by Elancheliyan et al., a rather high content of copolymerized acrylic acid suppresses the volume phase transition. On the contrary, the full NIPAM network interpenetrated with the acrylic acid one was found to maintain its thermo-responsiveness, as shown in Figure 4: the VPTT is shifted and, due to the presence of the weak acid, it is pH-dependent.

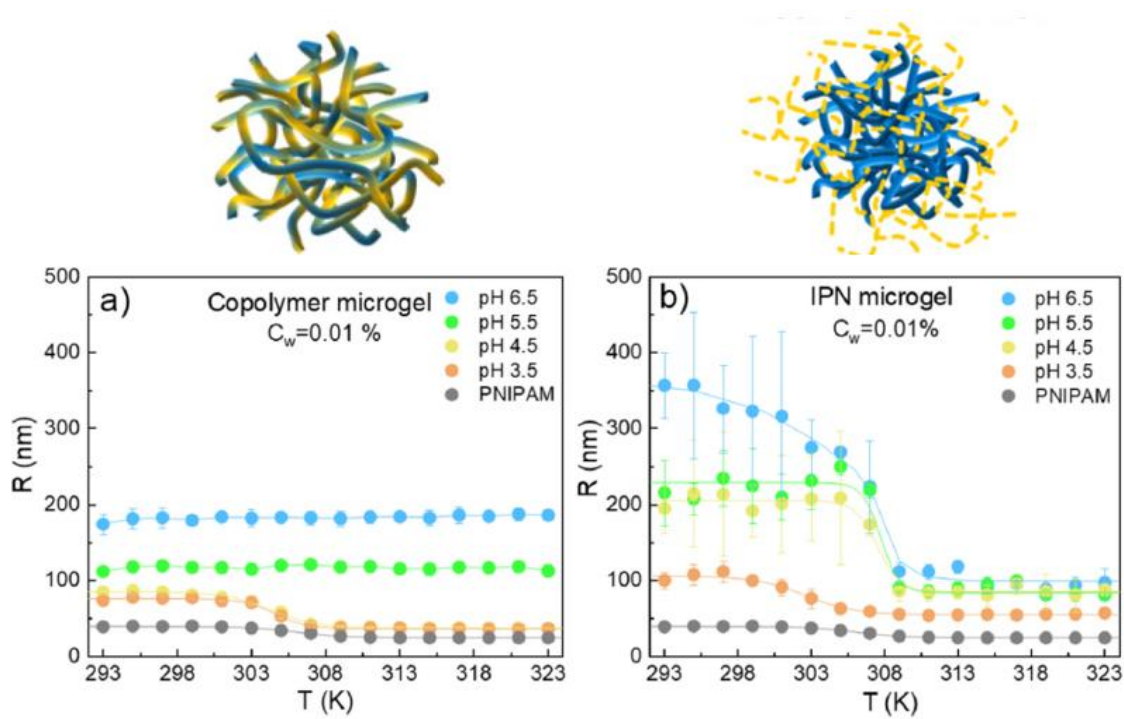

**Figure 4:** Hydrodynamic radii of PNIPAM microgels and P(NIPAM-co-AAC) copolymer microgels (a) and IPN microgels (b) obtained from DLS measurements as a function of temperature and pH in dilute conditions [74]. Lines are guides to the eye. The two structures are schematically illustrated as a visual reference Adapted from [74], licensed under CC BY 4.0.

At low pH, the particles are small and have a low VPTT reminiscent of the one of NIPAM, whereas at high pH, particles swell due to electrostatic repulsion, and the VPTT is shifted to higher temperatures by some 10 K. Similar results have been observed with a completely different technique, i.e. interfacial tension measurements, by Komarova et al. [75]: the interpenetrated networks allow the formation of more homogeneous particles, whereas copolymerization induces the formation of a shell of charged molecules, thus conferring adsorption properties which are different from the ones of the original NIPAM microgel. There is also an interesting parallel to the work by Cors et al. [36]: Both microgel systems are interpenetrated networks characterized by different VPTT. In the case of Buratti et al., one may imagine the thermosensitive PNIPAM-part of the copolymerized network to be constantly stretched by the charged, second part, thereby retarding and finally suppressing its collapse with increasing temperature. In the other system by Buratti et al., the independent NIPAM-network undergoes the VPT while being hindered by the second one, until both finally fully collapse, presumably due to the stronger impact of the hydrophobic pure NIPAM.

As we have seen, the presence of ions as part of the polymer network adds hydrophilicity and thus swelling. Another approach has been proposed by Ayazbayeva et al. aiming at a control of the overall electrostatic charge of the microgel by adding a mixture of comonomers of opposite charge to NIPAM, thus obtaining a zwitterionic microgel with balanced or unbalanced charge ratio [76]. Although there is no net charge for balanced microgels, there is an effect of ionic strength: an increase in the VPTT is observed, before losing colloidal stability at very high salt content. In presence of a total net charge, the volume phase transition behavior appears to be robust, and the VPTT remains high and approximately constant with (moderately) increasing salt.

The effect of copolymerization with methacrylic acid has also been studied by Sabadasch et al., with the aim of providing electrostatic binding sites for catalytic palladium nanoparticles to be included in the microgel [77]. The synthesis was performed at low pH, and the authors then investigated the difference in swelling properties at low and high pH. At pH 4, the methacrylic acid remains protonated, and the swelling curves showed a decrease in low-T particle size, as well as a progressive shift by some 5 °C between the pure PNNPAM microgel (20.9 °C, unchanged by pH) and the one with 25 mol% acid. At pH 10, with ionized molecules, the picture is completely different: the characteristic collapse was progressively replaced by a strongly widened transition, which is difficult to locate exactly but takes place at higher temperatures, above 40 °C. The catalytical activity has then been studied using a model reaction. Results show that while the nanoparticles are immobilized inside the microgels, the reaction of negatively charged molecules seems to be hindered by electrostatic repulsions caused by the introduced carboxylate comonomer.

Charged catalytic nanoparticles are not the only loading one may wish to deliver via temperature control with a microgel. Rodriguez-Tellez et al. have studied the release of an anti-cancer drug, doxorubicin by PNIPAM-PEGMA microgels copolymerized with the amphiphilic methacryloylamido hexanoic acid [78]. The microgels are found to be pH-responsive. At high pH, the acid group is deprotonated, and the resulting net charge of the microgel shifts its swelling curve to higher temperatures, and it broadens the volume phase transition. This is thus yet another example of a possible control of the VPTT by other comonomers. As far as the application is concerned, these authors show that the microgels have acceptable cytocompatibility, and that the drug can be loaded efficiently, and released under specific acidity and temperature conditions, namely at pH 5 at 42 °C, i.e. at temperatures which may be higher than the usual body temperature, for instance close to tumor cells.

## 6 Controlling thermosensitivity by copolymerization with ionizable comonomers, with focus on microgel morphology

In a symmetric way to section 3, we focus on articles highlighting the formation of original morphologies formed with ionizable comonomers in this last part. Again, we start with non-NIPAM systems, and namely a very early publication on acrylates [79], but also vinylcaprolactam-based systems, before discussing systems of the NIPAM family. Rodriguez et al. have pioneered the importance of pH-control of swelling via the introduction of deprotonatable comonomers of methacrylic acid into an ethylacrylate microgel [79]. By comparing the radius of gyration to the hydrodynamic one, they concluded on non-uniform swelling at high pH and low ionic strength, modeled by a denser crosslinked core, and a loser shell with dangling ends, and which is evidenced by a characteristic dip in the $R_g/R_H$ ratio. We have already seen in the last section that their findings are still relevant for today's publications [28] , and others will be discussed here.

Hussmann et al. have copolymerized vinylcaprolactam with a vinylimidazolium group in order to introduce controlled amounts of positive charges into the polymer network [80]*. Their contribution is noteworthy because the electrostatic charge introduced in most publications is negative, as it is created by deprotonation at high pH of weak (acrylic) acids. Whatever the sign of the charge, the effect of progressive charging is similar, as measured by the swelling curves: the VPTT shifts to higher temperatures, and the transition broadens. The motivation for the choice of this system is again the biocompatibility of VCL, and the possible interaction with E. Coli for antibacterial action. In this context, microgel morphology has its importance, and a core-shell analysis based on high resolution proton NMR shows that the charged groups participate in the collapse at higher temperatures than the VCL ones, and are located on the outside of the particles, where possible binding to E.Coli can take place. Indeed, confocal light microscopy shows clustering of cells with increasing charge content, which is correlated with cell death. The work by Hussmann can be put in parallel to the one by Harsanyi et al., who overcome the limitations of incorporation by copolymerization by cationic surface-modification of PNIPAM microgels, also in view of biomedical applications, like transfection, or RNA or enzyme packing [81].

The effect of copolymerization with chargeable comonomers has been extensively studied for NIPAM-based microgels. In their recent article, Hazra et al. continue work on large, micron-sized hollow microgels, introducing charged groups [82]*. This allows a direct comparison to the previous, non-charged system [83]. One may note that more complex architectures, in particular hollow double-shell systems, have been proposed and studied by the same group, as e.g. by Brugnoni et al. [84]. Here we concentrate on the comparison of the simple hollow shell in the charged and uncharged state. The hollowness imparts a very high softness to the particle, as probed by osmotic compression and evidenced by buckling. The synthesis route chosen by this group is based on the use of a sacrificial silica core, on which a NIPAM-itaconic acid microgel has been copolymerized. The acid groups provide responsiveness to salt and pH, making the particle swelling sensitive to multiple experimental parameters. In particular, it is found that at low pH, the usual thermosensitivity is found, whereas in the charged state, the electrostatic repulsion impedes any high temperature collapse. In parallel, buckling is observed for virtually all particles at low pH under osmotic stress set by the addition of 10 wt % Dextran, whereas a high fraction of particles seems to be undeformable at high pH in the same osmotic conditions.

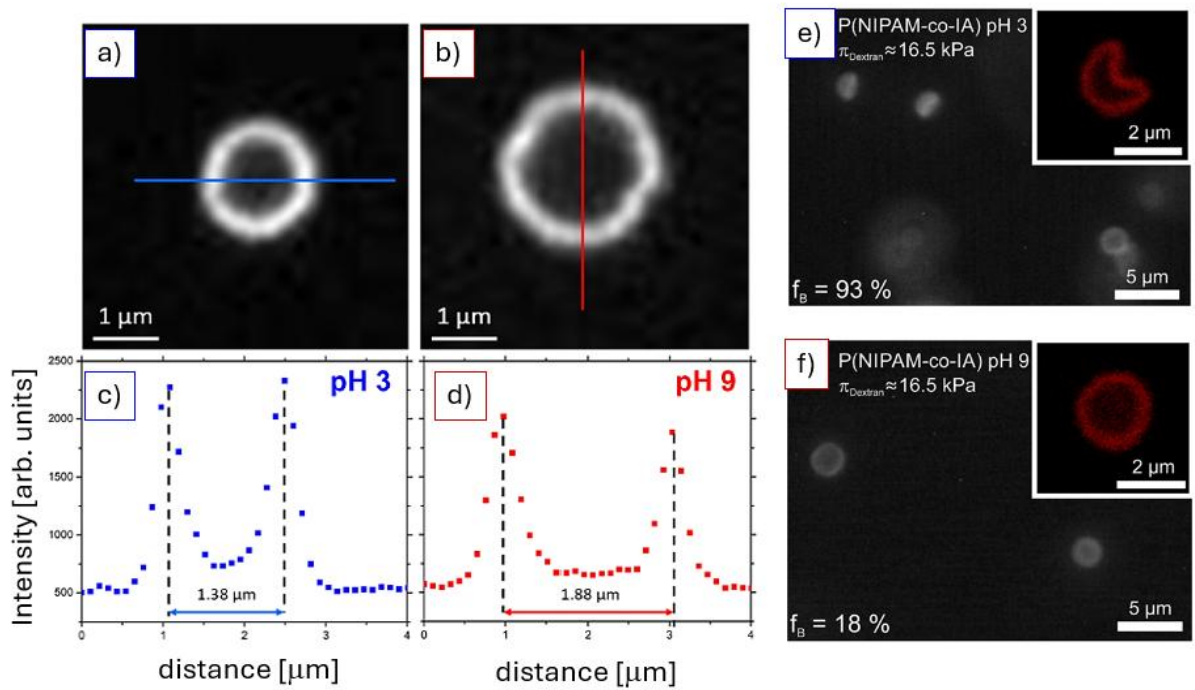

**Figure 5:** a-b) Confocal spinning disk fluorescence microscopy (SDFM) images (top) and c-d) associated intensity profiles of P(NIPAM-co-IA) capsules in aqueous solution at pH = 3, I = 10 mM and pH = 9, I = 10 mM. e-f) Microgel capsules under osmotic stress set by the addition of 10 wt% Dextran (pDextran E 16.5 KPa). Fluorescence micrographs are shown for charged microgels capsules at pH = 3 (e) and charged microgel capsules at pH = 9 (f). The insets show the higher magnification of the capsules recorded via confocal laser scanning microscopy (CLSM). The proportion of buckled capsules, $f_B$, was determined by statistical analysis of the fluorescence micrographs. Adapted from [82]*, licensed under CC BY-NC 3.0.

Surface functionalization of PNIPAM microgel cores has been proposed by Guerron et al. as a means of modifying adsorption properties without impacting the thermosensitivity of the core [85]. By grafting either neutral PEG or charged, pH-sensitive amine groups (PDMAEMA), they introduced a moderate pH-sensitivity independently of the swelling of the core. The latter was shown to be close to the expected swelling curves of pure NIPAM for both surface modifications.

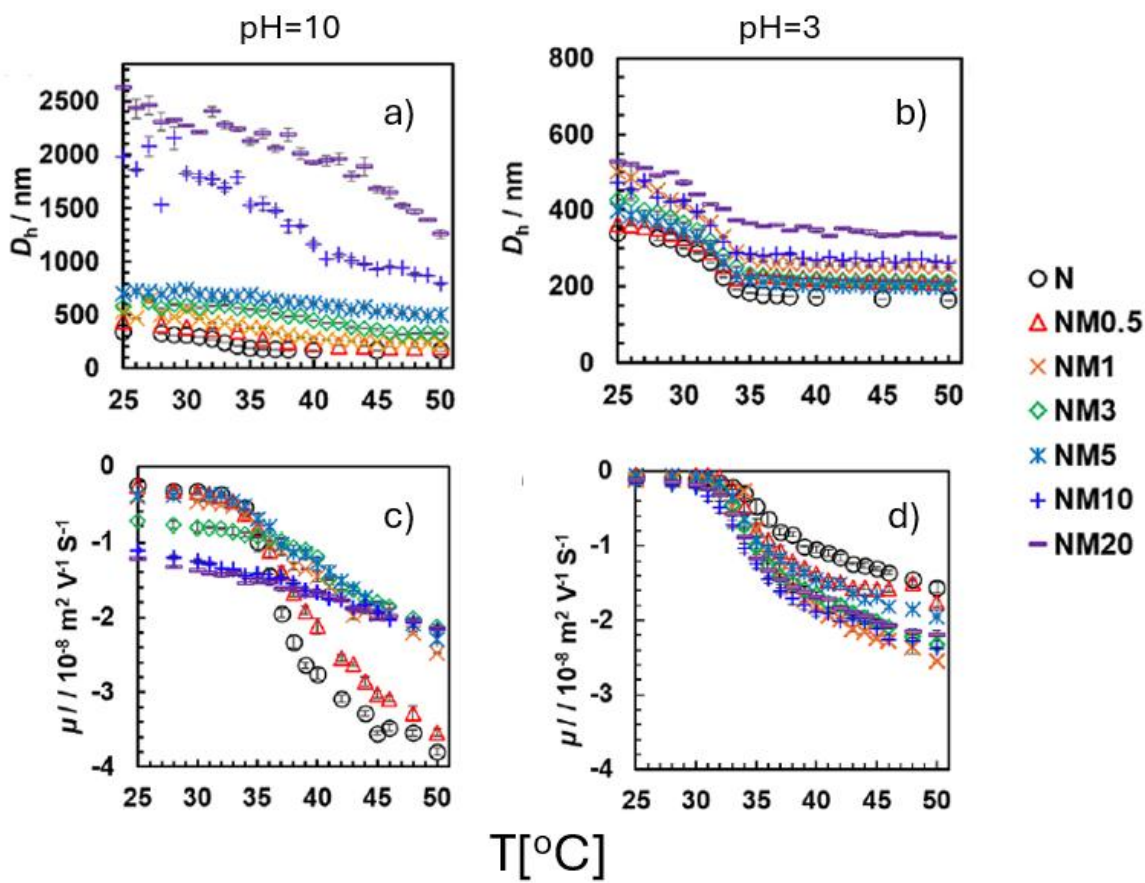

**Figure 6:** Temperature dependence of the hydrodynamic diameter $D_h$(a,b) and electrophoretic mobility µ (c,d) at pH 10 and pH 3 of NM*x* copolymer microgels studied in [86]. Adapted with permission from [86], Copyright 2025 American Chemical Society.

The generation, control, and effect of monomeric gradients of charged carboxyl groups on the swelling and the surface properties of microgels has also captured the attention of Nishizawa et al. [86]. They compare NIPAM copolymers made with methacrylic acid (MAc) (Figure 6) to others also copolymerized with fumaric acid. The effect of methacrylic acid is in line with previously discussed literature: for charged microgels at high pH, the volume phase transition broadens, shifts to higher temperature, and finally disappears.

More specifically, at pH = 10, the hydrodynamic diameter of the NIPAM-co-methacrylic acid (NM*x*) microgels first gradually and then drastically decreased during heating from 25 to 50 °C (Figure 6 a). Here *x* stands for the molar fraction of the methacrylic acid comonomer. It should also be noted that NM10 and NM20 swelled significantly, and that the deswelling transition was not completed, not even upon heating to 50 °C. This recalls the very similar jump in size observed for a rather different copolymerization performed by [64]. Interestingly the absolute values of the electrophoretic mobility *μ* showed a remarkable increase with increasing temperature at pH = 10, pointing to an increase in polymer density via deswelling (Figure 6c). In the presence of large amounts of MAc (>10 mol %), drastic changes in *μ* were not observed. At low pH (pH 3, Figure 6 b-d), the VPT remains virtually identical to the one of the PNIPAM microgel, with an overall reduced effect of methacrylic acid content on the deswelling and electrophoretic behavior upon heating. The authors further conclude that the methacrylic acid is preferentially incorporated in the core, without much effect on electrokinetic surface properties. This gradient structure leads to a multi-step collapse, as outer regions still have a VPTT reminiscent of the one of PNIPAM. In additional presence of fumaric acid, however, the lower reactivity of the latter leads to an enrichment in fumaric acid of the peripheral zones, with corresponding changes in surface properties.

## 7 Conclusion

We have reviewed recent contributions to the synthesis and study of the swelling behavior of microgel particles made by copolymerization with comonomers allowing changes in the VPT, sometimes associated with non-trivial internal structures such as gradient, core-shell, multilobular, or interpenetrated polymer networks. Modifying the hydrophobic/hydrophilic balance of microgels by introducing new monomers is a straightforward way of shifting the VPT, as long as they are hydrophobic enough to allow precipitation polymerization to take place at accessible temperatures in water – usually some 70 °C, but sometimes approaching 100 °C. The differences in hydrophilicity of the monomers changes the concentrations of each species in the nucleation and growth steps of the synthesis, and therefore may induce the formation of heterogeneous, non-trivial final microgel structures, with correspondingly novel swelling properties. As a general observation, it seems that the more original structures are generated by the weaker (hydrophobic) forces, which give more space for reorganization. On the other extreme, strong electrostatic repulsion always induces additional swelling, and possibly molecular gradients.

In some cases, the different monomer reactivities were used on purpose to generate substructures of different thermosensitivity, whereas other contributions aimed at using new structures, such as interpenetrated networks, to achieve a high degree of molecular homogeneity. As shown in this review, the resulting multi-functionality may be used for specific applications, like drug release from one compartment, or film formation. The same is true when introducing comonomers with additional functional groups, including electrostatic charges, or additional light-sensitive moieties capable of photo-isomeration. In the latter case, the molecular conformations can be reversibly switched between different hydrophobicities by light irradiation, providing a clever external trigger.

One may think about alternative ways to modulate the swelling capacity of microgel particles, other than copolymerization of (small) monomers. An interesting pathway has been proposed for macroscopic hydrogels by Kalkan and Orakdogen [87]. They have studied the thermosensitive swelling properties of NIPAM hydrogels not only in the presence of charge, but also with linear polyacrylamide incorporated into the gel, producing veritable IPN microgels. This idea is related to introducing macromonomers [88,89] in a microgel, thereby possibly introducing (zwitter)ions, changing swelling, and as an application, protein-repelling antifouling properties. Alternatively, one may also inspire oneself by other ways of crosslinking the microgels. Karga et al., for instance, were motivated by shear thinning properties introduced by physical, and thus self-healing crosslinks, as opposed to strong chemical crosslinks formed by BIS [90]. By incorporating hydrophobic stickers in their hydrogels – the presence of individual microgels remains to be proven –, they succeed in creating such injectable, self-healing gels. One may wonder if a similar mechanism would allow rearrangements of network connectivity under swelling in microgel particles.

To summarize, we have shown that fine-tuning the physical, chemical, and biological properties of microgel particles is receiving considerable attention given the rather mature field of studies. We have chosen to discuss and highlight articles in this review that appeared to us to be the most striking illustrations of a much larger set of publications. We hope to have convinced the reader that many groups have succeeded in generating multi-responsive (pH, salt, light) structures, sometimes bearing multi-compartment morphologies, the different thermosensitivities of which can be individually addressed. Indeed, selectively tuning their solvation state provides a key to size control – and thus solvent permeability, molecular accessibility, softness, etc. – on the scale of subregions of microgels, with possible mechanical, optical, catalytic, chemical, or electrostatic effects on their surroundings.

## Declaration of competing interest

There are no competing interests to disclose.

## Acknowledgements

J.O, T.H. and D.T. are grateful for funding of the joint 'SmartBrane' project by the DFG (Grant HE 2995/14-1) and the ANR (Grant ANR-22-CE92-0052-01). D.T acknowledge financial support from the Agence Nationale de la Recherche (Grant ANR-20-CE06-0030-01; THELECTRA).

**This study on the effect of large ions binding to neutral polymer on the swelling capacity of microgels is remarkable because it also includes a structural study by small angle scattering of light (SLS) and X-rays (SAXS). In particular, the adaptation of the particle geometry to ions is shown to depend on network topology controlled via the degree of crosslinking of the microgel.**

**This interesting study combines an approach in synthetic chemistry – copolymerize NIPAM with methylcellulose – with a rather involved study of loading and release of Genipin, including its effect on cells.**

**This study aims at connecting microgel particles into a polymer film by copolymerizing a film-forming molecule of low T$_g$ in a second process. We emphasize this contribution because of the peculiar morphology (lobes), which can be controlled via the kinetics and thus the temperature, and which has moreover been described by a model. Also, film swelling is investigated, and moreover films are employed here for protein release.**

**The contribution by Gruber et al. proposes an impressive synthetic platform for multifunctional microgels which have an amphiphilic interior for loading hydrophobic guest molecules, while being stimuli-responsive and possibly pH-degradable for controlled release. They add a functionalizable hydrophilic shell to minimize aggregation and enable controlled interactions with biological systems .**

**The dominant influence of the crosslinker on the network architecture is discussed in this article, based on a combination of small-angle scattering modelled in detail, and rheology of dense suspensions. Moreover, microgels are observed under osmotic compression by PEG invisible to neutrons, showing their progressive deswelling, opening the road to a comparison between temperature and pressure induced deswelling.**

**In this work, azobenzene-functionalized light, temperature, and pH multiresponsive microgels were successfully fabricated. The authors convincingly showed that the synthesized microgels exhibit reversible photoswitching behavior upon blue-, green- or red-light irradiation because of the trans−cis isomerization induced by different visible light irradiations. In addition to that, the net charge of the particles could be tuned by adjusting the pH.**

**Contrary to mixing chemically different comonomers within the chains outlined throughout this review, the influence of the network topology on the swelling properties of microgel particles seems to be one of the keys to tailoring particles for specific applications. Moreover, the combination of different techniques correlates in an impressive way molecular insight on hydration, eg. via Raman spectroscopy of $CH_3$ stretching, with the swelling observed by DLS.**

**Besides the particularity of incorporating positive charges into a microgel, this article convinces with a precise measurement of microgel deswelling by NMR using proton transverse magnetization relaxation. This technique is capable of resolving spatially, via a subdivision into core and corona regions, where the charged groups are located, through their effect on methyl(ene) group dynamics. This is then successfully correlated by imaging with interaction with bacteria.**

**The formation of complex architectures, such as empty-core microgel particles, opens a fascinating field of study in itself, especially with respect to concepts of compression, softness, and potentially rheological response. Here the authors present a multi-tunable hollow shell, with a striking visualization of tunable deformability in response to osmotic compression.**